\documentclass[10pt,conference]{IEEEtran}
\usepackage{booktabs}
\usepackage{array}
\usepackage{graphicx}
\usepackage{amsmath}
\usepackage{amssymb}
\usepackage[table]{xcolor}
\usepackage[normalem]{ulem}  
\usepackage[numbers]{natbib}
\usepackage{xfrac}
\usepackage{braket}
\usepackage{subcaption}
\usepackage{comment}
\usepackage{float}
\usepackage{mathtools}
\usepackage{tikz}
\usepackage[]{mdframed}
\usepackage{tabularx}
\usepackage{multirow}
\usepackage{algorithm}
\usepackage{algpseudocode}
\usepackage{soul}
\usepackage[colorlinks=true,allcolors=.]{hyperref} 

\definecolor{green}{rgb}{0.2, 0.9, 0.0}

\newcommand{\alexeyshift}{4.0em}

\DeclareMathOperator*{\argmax}{arg\,max}

\begin{document}
%

\title{Pauli Correlation Encoding for mRNA Secondary Structure Prediction: Problem-Aware Decoding for Dense-Constraint QUBOs}

\author{
\IEEEauthorblockN{Triet Friedhoff}
\IEEEauthorblockA{\textit{IBM Quantum}\\
New York, USA \\
triet.nguyen-beck@ibm.com}
\and
\IEEEauthorblockN{Mihir Metkar}
\IEEEauthorblockA{\textit{Moderna}\\
Cambridge, USA \\
mihir.metkar@modernatx.com}
\and
\IEEEauthorblockN{Wade Davis}
\IEEEauthorblockA{\textit{Moderna}\\
Cambridge, USA \\
wade.davis@modernatx.com}
\and
\IEEEauthorblockN{Vaibhaw Kumar}
\IEEEauthorblockA{\textit{IBM Quantum}\\
New York, USA \\
vaibhaw.kumar@ibm.com}
\and
\IEEEauthorblockN{\qquad\qquad\qquad\qquad\qquad\hspace*{\alexeyshift}}
\and
\IEEEauthorblockN{Alexey Galda}
\IEEEauthorblockA{\textit{Moderna}\\
Cambridge, USA \\
alexey.galda@modernatx.com}
\and
\IEEEauthorblockN{\qquad\qquad\qquad\qquad\qquad}
}

\bstctlcite{IEEEexample:BSTcontrol}
\maketitle

\begin{abstract}
Pauli Correlation Encoding (PCE) compresses $m$ binary variables onto $n = \mathcal{O}(m^{1/k})$ qubits for a tunable compression order $k \geq 2$ by mapping them to commuting Pauli correlators, but the continuous expectation values it produces must be decoded into feasible binary solutions, a challenge that becomes acute for problems with dense constraints. We apply PCE to the mRNA secondary structure prediction problem, formulated as a densely-constrained QUBO. We train throughout with a QUBO-space sigmoid loss that preserves the QUBO penalty structure directly. For decoding, we introduce the Problem-Aware Guided Decoder (PAGD), which scores candidate variable commitments by the product of their marginal QUBO energy reduction and a trained expectation-value prior, with constraint-aware feasibility pruning at each step. On six benchmark mRNA sequences (30--60~nucleotides (nt), 50--240 variables, 7--14 qubits), PAGD with 100 restarts (PAGD-$K_{100}$) achieves 75--100\% near-optimal recovery, defined as P(gap$<$1\%), for sequences up to $m{=}152$ variables, vs.\ 0--30\% for the sign-rounding+local-search baseline (Sign+LS). We demonstrate that PCE-trained expectation values provide a useful prior: on \texttt{seq\_240}, trained PAGD improves with restart count and reaches 50\% P(gap$<$1\%) at $K{=}200$, outperforming the untrained-circuit control by 10 percentage points and the random-EV baseline by 40 percentage points. Hardware-scale results extend the pipeline to three mRNA sequences of 102--105 nucleotides ($m{=}694$--$745$ variables, $172{,}000$--$193{,}000$ pair constraints, $n{=}23$ qubits) deployed on IBM Heron processors as single 100-iteration QPU runs per sequence. The variational circuits transpile SWAP-free into $480$ native two-qubit gates at depth $256$, and the PAGD-$K_{200}$ decoded gap on the QPU matches or beats the simulator mean for all three sequences. \textbf{Notably, a single 100-iteration QPU run on the 102-nucleotide sequence recovers the CPLEX optimum \emph{exactly} (0.0\% gap) at $K{=}200$ and reaches a $7.4$\% gap already at $K{=}10$. The result is consistent with the matched simulator distribution and demonstrates that the PCE-trained prior can survive transit to noisy superconducting hardware at biologically relevant scale.}
\end{abstract}

\IEEEpeerreviewmaketitle

\section{Introduction}\label{sec:introduction}

Predicting the minimum free energy (MFE) secondary structure of mRNA is a key step in mRNA-based therapeutic design: the folding pattern can determine translation efficiency, immunogenicity, and overall stability of the molecule. While dynamic-programming algorithms solve the MFE problem exactly for short sequences, the combinatorial complexity introduced by pseudoknots makes the problem NP-hard for sequences of therapeutic relevance, which span hundreds to several thousand nucleotides~\cite{lyngso2000pseudoknots}. Recent work has demonstrated that quantum optimization can address this problem by formulating it as a quadratic unconstrained binary optimization (QUBO) and solving it variationally on superconducting processors~\cite{alevras2024mrna, kumar2025longer}. Like prior quantum work, we address the non-pseudoknotted case. The largest such demonstration to date scaled to 354 qubits in tensor-network simulation, and deployed circuits with up to 156 qubits and 950 nonlocal gates on IBM Heron processors for sequences of 60 nucleotides~\cite{kumar2025longer}. However, these approaches map each binary variable to a dedicated qubit, and the qubit count (equal to the number of QUBO variables in these one-variable-per-qubit formulations) grows approximately as $L^{2.2}$, where $L$ is the number of nucleotides~\cite{kumar2025longer}. A more qubit-efficient encoding is therefore needed to bring quantum approaches closer to therapeutically relevant problem sizes.

Pauli Correlation Encoding (PCE) was introduced as a qubit-efficient variational framework for combinatorial optimization~\cite{sciorilli2025pce}. By mapping multiple binary variables onto distinct commuting Pauli correlators (XX, YY, ZZ, or subsets thereof) measured on the same small qubit register, PCE achieves polynomial compression: \(\mathcal{O}(n^k)\) variables can be encoded using only \(n\) qubits for a tunable compression order \(k \geq 2\). The original paper~\cite{sciorilli2025pce} demonstrated that this encoding, paired with a shallow variational circuit and a simple sign-based decoder, suppresses barren plateaus and produces competitive approximation ratios on Max-Cut instances with thousands of variables, all while remaining executable on near-term hardware.

Subsequent work extended PCE to portfolio optimization over $250{+}$ assets via iterative graph partitioning~\cite{soloviev2025portfolio} and to budget-constrained Min-Cut and TSP~\cite{padin2026budget}.

Despite this growth, decoding remains the open question. PCE produces a relaxed fractional solution that must be rounded to a binary one; on dense-constraint QUBOs the relaxed and binary optima can diverge, and naive sign rounding~\cite{sciorilli2025pce} ignores the inter-variable correlations the encoding induces, frequently producing infeasible solutions and forcing ad-hoc loss regularizers~\cite{sciorilli2025competitive} that mask the underlying issue.

Ansatz design is similarly under-explored. The original problem-agnostic brickwork ansatz~\cite{sciorilli2025pce} has gate and parameter counts that scale linearly with $m$, large enough that classical optimizers struggle; a QUBO-informed ansatz is a natural way to retain expressivity without inflating parameters.

We present a systematic investigation of the training--decoding pipeline for PCE on mRNA secondary structure prediction, characterized by dense Maximum Independent Set (MIS)-like constraints. Our contributions are: (i) a problem-informed ansatz (Informed-$k$) that ranks qubit pairs by QUBO importance and connects them via maximum spanning tree, outperforming nearest-neighbor and denser topologies on most tested sequences; (ii) the Problem-Aware Guided Decoder (PAGD), which scores variable commitments by the product of marginal QUBO energy reduction and the trained expectation-value (EV) prior, with constraint-aware feasibility pruning; (iii) demonstration that PCE-trained EVs provide a useful prior, with trained PAGD on \texttt{seq\_240} reaching 50\% P(gap$<$1\%) at $K{=}200$ and outperforming the untrained-circuit and random-EV controls by 10 and 40 percentage points, respectively; and (iv) hardware-scale results on three $\sim$100-nt instances (\texttt{seq\_694}, \texttt{seq\_715}, \texttt{seq\_745}; $m{=}694$--$745$, $n{=}23$) on IBM Heron quantum processors (QPUs) \texttt{ibm\_pittsburgh} and \texttt{ibm\_aachen}. We adopt the QUBO-space sigmoid loss throughout as the principled default aligned with the decoder. We verified empirically that on these instances solution quality under the QUBO-space loss is comparable to the Ising-space loss; the full head-to-head comparison falls outside the scope of this paper and is omitted for space.


\section{QUBO Formalism}\label{sec:QUBO}
Each position in an mRNA sequence ($i = 1,\ldots,L$) carries one of the bases $\{U, A, C, G\}$. A base pair $(i, j)$ is valid if the two positions form one of $\{AU, UA, CG, GC, GU, UG\}$, and the QUBO decision variable is a \emph{quartet} (stacked pair) $(i, j, i+1, j-1)$ comprising two consecutive base pairs.

Following the framework established in Ref.~\cite{gusfield2019integer}, we define the quartet (stacked pair) variable as:

$$x(i,j,i+1,j-1)=\left \{
\begin{array}{rl}
1 &\text{if stacking occurs between base}\\
& \text{pairs $(i,j)$ and $(i+1,j-1)$} \\
0 &\text{otherwise.}
\end{array}
\right .
$$ 
Variables exist only for valid base pairs. Two physical constraints apply: each base pairs with at most one other, and base pairs may not cross:
\begin{equation} \label{eq:mis}
x(i,j,i+1,j-1) + x(k,\ell,k+1,\ell-1) \le 1.   
\end{equation}
This type of constraint is also observed in the independent set problem~\cite{glover2018tutorial}. Defining 
$q_i$ as the variable for the $i$-th quartet and denoting $Q$ as the set of all valid quartets, $QC$ as the set of crossing quartet pairs, $QS(q_i)$ as the set of quartets that can be stacked with $q_i$, and $\mathit{QUA}$ as the set of stacked quartets ending in a $(UA)$ pair, we express the corresponding QUBO problem as follows:
\begin{align}\label{eq:QUBO} 
     \min &\sum_{q_i \in Q} e_{q_i} q_i  +
r\sum_{q_i \in Q}{\sum_{q_j \in QS(q_i)}}q_i q_j \nonumber \\  
& + p \sum_{q_i \in Q} \sum_{q_j \in \mathit{QUA}} q_i (1-q_j) + t \sum_{q_i,q_j \in QC} q_i q_j. 
\end{align}

Each quartet $q_i$ has a free energy $e_{q_i}$ from empirical thermodynamic data~\cite{turner-mathews}; $r$ rewards consecutive (stacked) quartets~\cite{turner-mathews}; $p$ penalizes structures ending in a less stable $(UA)$ pair; and $t$ enforces the no-crossing constraint as a penalty.

These QUBO problems typically exhibit high edge densities (0.7--0.85) in their constraint graphs. The resulting large penalty coefficients dominate the energy landscape and make it difficult to distinguish feasible from infeasible solutions on the relaxed objective alone, motivating the constraint-aware decoder and problem-informed ansatz introduced next.

\section{Methods}\label{sec:methods}

\begin{figure*}[t]
    \centering
    \includegraphics[width=1.0\linewidth]{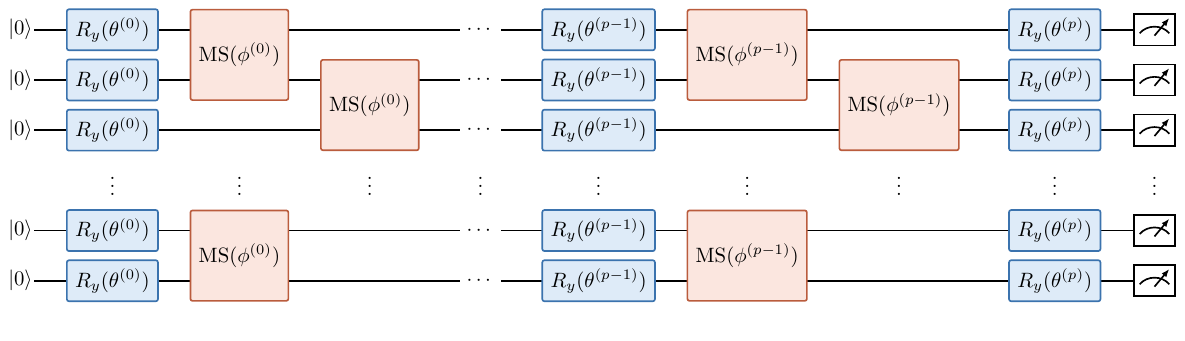}
    \caption{Circuit ansatz under the \emph{layerwise} parameterization adopted throughout this work: a single $R_y$ angle $\theta^{(\ell)}$ is shared across all qubits within layer $\ell$, and a single MS angle $\phi^{(\ell)}$ is shared across all entangling pairs in layer $\ell$, yielding $2p+1$ trainable parameters at depth $p$. The MS-gate connectivity shown is the nearest-neighbor (NN) topology, i.e.\ a Hamiltonian path through the $n$ qubits drawn here in the standard brickwork layout (alternating non-overlapping sublayers); the actual qubit pairs depend on the chosen ansatz topology (NN, Informed-$k$, Informed-$2k$, or All; see Sec.~\ref{sec:informed-ansatz}).}
    \label{fig:ansatz}
\end{figure*}

\subsection{Pauli Correlation Encoding}
\label{sec:pauli-correlation}
PCE maps $m$ binary QUBO variables onto $n$ physical qubits via two-body Pauli correlations. Each $x_k$ is encoded as the expectation value of $\hat{P}_{q_i q_j} \in \{X_iX_j, Y_iY_j, Z_iZ_j\}$ on a designated qubit pair, giving a capacity of $3\binom{n}{2}$. All three correlators can be estimated in three measurement settings on a QPU~\cite{sciorilli2025pce}. The minimum qubit count $n^*$ satisfies
\begin{equation}
    3\binom{n}{2} \;\geq\; m.
    \label{eq:min-qubits}
\end{equation}

\subsection{Problem-Informed Ansatz Construction}
\label{sec:informed-ansatz}

We use a variational ansatz similar to previously published PCE-based papers~\cite{sciorilli2025pce, sciorilli2025competitive, padin2026budget}.
The variational ansatz, shown in Fig.~\ref{fig:ansatz}, follows a layered structure consisting of $p$ repetitions of alternating single-qubit rotation and two-qubit entangling layers, followed by a final rotation layer:
\begin{equation}
    U(\boldsymbol{\theta}) \!=\! \Bigl[\prod_{q} R_y\!\bigl(\theta_q^{(p)}\bigr)\Bigr] \prod_{\ell=0}^{p-1} \Bigl[ \mathcal{E}_\ell \prod_{q} R_y\!\bigl(\theta_q^{(\ell)}\bigr) \Bigr],
    \label{eq:ansatz}
\end{equation}
where $R_y(\theta) = e^{-i\theta Y/2}$ and $\mathcal{E}_\ell$ is the entangling layer at depth~$\ell$. Each entangling layer applies parameterized M{\o}lmer--S{\o}rensen (MS) gates, $\mathrm{MS}(\theta) = \exp\!\Bigl[\!-i\tfrac{\theta}{4}(XX + YY)\Bigr]$ on a specified set of qubit pairs~$\mathcal{P}$.

Note, this gate is native to the trapped-ion quantum platform and requires transpilation to the basis gate set on IBM superconducting devices~\cite{abughanem2025hardware}. We adopt the \emph{layerwise} parameterization throughout: a single $R_y$ angle and a single MS angle per layer are shared across all qubits and all pairs in that layer, yielding $2p + 1$ trainable parameters in total. This drastically reduces the optimizer's burden compared to assigning an independent parameter to every gate, which we found essential for the Constrained Optimization BY Linear Approximation (COBYLA) algorithm~\cite{powell1994cobyla} to converge at the circuit depths used here. We use COBYLA throughout because the layerwise parameterization gives a small ($\le 41$ for our largest hardware instance) parameter space where COBYLA's trust-region updates converge in tens to hundreds of iterations without requiring gradient estimates. We also benchmarked Nelder--Mead~\cite{nelder1965simplex}, Powell's method~\cite{powell1964efficient}, L-BFGS-B~\cite{byrd1995lbfgsb}, and SPSA~\cite{spall1992spsa} at matched wall-clock on representative instances; COBYLA was the most accurate and most consistent across seeds, and we adopt it as the default optimizer throughout this work.

\paragraph{\textit{Ansatz topologies}}
We benchmark four entangling topologies, keeping the same circuit depth $p$ and parameterization scheme across all variants to ensure a fair comparison:

\begin{itemize}
    \item \emph{Nearest-neighbor (NN).} $k = n{-}1$ pairs forming a Hamiltonian path through the $n$ qubits, with qubit ordering randomized per seed. Problem-agnostic, so entangling capacity may be wasted on pairs that carry little QUBO weight.

    \item \emph{Informed-$k$.} $k = n{-}1$ pairs selected by QUBO importance (defined below) via a Kruskal maximum-spanning-tree pass~\cite{kruskal1956spanning}: candidates are added in descending importance order and accepted only when they join two disconnected components, yielding a connected spanning tree. The pair set is deterministic across seeds; only the optimizer initialization varies.

    \item \emph{Informed-$2k$.} Same QUBO-importance ranking but with a doubled budget $k = 2(n{-}1)$: the first $n{-}1$ pairs come from the spanning-tree pass; the remaining $n{-}1$ are the next-highest-importance unused pairs. Deterministic across seeds.

    \item \emph{Fully connected (All).} All $\binom{n}{2}$ pairs included per layer, with gate ordering randomized per seed (MS gates on different pairs do not commute). Serves as an upper bound on connectivity at the cost of the largest gate count.
\end{itemize}

\begin{table*}[t]                                                                                                                                                                  
  \centering                                                                                                                                                                         
  \caption{Benchmark sequences used in Figs.~\ref{fig:ansatz_topology}--\ref{fig:decoder_comparison} (Informed-$k$ ansatz topology, layerwise parameterization), together with the
  three hardware-scale instances reported in Sec.~\ref{sec:hw-scale}. $m$: QUBO variables; $L$: nucleotides; $n$: minimum qubit count satisfying $3\binom{n}{2}\geq m$; $|QC|$:      
  number of pair (no-crossing) constraints in the QUBO conflict graph. Under the \emph{layerwise} scheme the COBYLA parameter count is $2p{+}1$, independent of $n$ and of the
  topology.}                                                                                                                                                                         
  \label{tab:benchmarks}                 
  \small                                         
  \setlength{\tabcolsep}{4pt}
  \renewcommand{\arraystretch}{1.15}                                                                                                                                                 
  \begin{tabularx}{\textwidth}{l r r r r >{\footnotesize\raggedright\arraybackslash}X}
  \toprule                                                                                                                                                                           
  ID & $m$ & $L$ & $n$ & $|QC|$ & \normalfont Sequence \\
  \midrule                                                                                                                                                                           
  \texttt{seq\_50}  &  50 &  30 &  7 &     991 & AAGCCUAUCAACGGCGUGCGCUGUGAUAUG \\
  \texttt{seq\_80}  &  80 &  42 &  8 &   2{,}345 & CCAUUUAUACCCCGGGCCGCCUACUGCACCCAGUGUAACAUG \\                                                                                     
  \texttt{seq\_120} & 120 &  45 & 10 &   5{,}427 & AUCCAUCAGUGGUACUGCAUGAUGCCCAUCUGCAUGGCCAAGAGG \\                                                                                  
  \texttt{seq\_152} & 152 &  60 & 11 &   8{,}236 & AUGAACCCCGACUGGACCCUGGCCAGAGUGAAGAGAAUCAUCAUCCUGCCCCACGAGUUC \\                                                                   
  \texttt{seq\_195} & 195 &  60 & 12 &  13{,}949 & GCUCGCGAAGAUCAGGGAUCACGAUGCUCGAAUUUUUAGACGAUCUAUAACGCCUCUCGG \\                                                                   
  \texttt{seq\_240} & 240 &  60 & 14 &  21{,}050 & GUCGGCAGCCUCGUGCCACCGCGAAGCGGUGCGAGCGCCGUGUCCAUGCCUACGGAGAUC \\                                                                   
  \texttt{seq\_694} & 694 & 102 & 23 & 172{,}307 & UCAACCGAAGGAGAAUCUUCCUUAAGAGCACCACGUAAGUGGGCCGGAUUUAGAUCUCCU\newline UAUAAGAAAGGGCUGCAUUCAGGUGCGCUUGUUUGCCAGCCC \\                
  \texttt{seq\_715} & 715 & 105 & 23 & 182{,}633 & AAAAGGGAUCUUGCGAUCAAUUAGUAUACCCUUAAUCCUGAUCGUUGCUUCACAACCUAA\newline AUGAAGUCUUAGUUGGACAAUCGAGUUUCACGGGUCGCUAAUAGA \\             
  \texttt{seq\_745} & 745 & 105 & 23 & 192{,}959 & UUCUCUAUUCUGAAUAGCCCUCCAGUCACCUCUUAAGAAUCGGUCAGGUACAUUAGAAAG\newline CAAAGCAAAAGGAUUUUUAGAGCGUUAUGGACAUUCAUUGGGUAC \\             
  \bottomrule                                                                                                                                                                        
  \end{tabularx}                                                                                                                                                                      
  \end{table*} 

\paragraph{\textit{QUBO importance score}}

For every QUBO coupling $(i,j)$ with $w_{ij}{=}|Q_{ij}|+|Q_{ji}|$, the importance accumulates two contributions: full weight $w_{ij}$ at the qubit pairs that \emph{encode} $i$ and $j$ (the \emph{direct} pairs), and half weight $w_{ij}/2$ at the up-to-four \emph{cross} pairs spanning the encodings of $i$ and $j$. The half weight reflects that cross pairs build correlations indirectly via the ansatz rather than encoding either variable directly:
\begin{multline}
    I(q_a, q_b) = \!\!\sum_{\substack{i < j \\ w_{ij} > 0}} \!\! \Big[ w_{ij} \bigl(\mathbb{1}_{\mathrm{pair}(i)} + \mathbb{1}_{\mathrm{pair}(j)}\bigr) \\
    + \tfrac{1}{2}\, w_{ij} \cdot \mathbb{1}_{\mathrm{cross}(i,j)} \Big],
    \label{eq:pair-importance}
\end{multline}
where $\mathrm{cross}(i,j)$ denotes the up to four cross-qubit pairs and the indicators test membership of $(q_a, q_b)$. A single qubit pair can accumulate contributions from many QUBO couplings, since multiple variables may share the same qubit. The top-$k$ pairs by $I$ form the ansatz connectivity; for $k > n{-}1$ the next-highest-importance pairs fill the remaining budget. We use $k = n{-}1$ throughout, so $\mathcal{P}^* = \mathrm{MST}_{\max}(I)$.

\paragraph{\textit{Hardware-aware pair selection}}
The topology-agnostic Informed-$k$ selection above does not enforce that selected qubit pairs are native edges of the target QPU. On IBM Heron devices, requiring a non-native entangling gate costs at least one SWAP per layer per pair, quickly inflating circuit depth beyond what is feasible with current gate fidelities. We therefore use a hardware-aware variant for QPU deployment: given the native edge set $\mathcal{E}_G$ of the best-connected $n$-qubit subgraph of the device~\cite{chamberland2020heavyhex}, each candidate pair receives a SWAP-discounted importance score
\begin{equation}
\begin{aligned}
  \tilde{I}(q_a, q_b) &= \frac{I(q_a, q_b)}{1 + \lambda\, \sigma(q_a, q_b)}, \\
  \sigma(q_a, q_b) &= \max\bigl(d_G(q_a, q_b) - 1,\, 0\bigr),
\end{aligned}
  \label{eq:hw-aware}
\end{equation}
where $d_G$ is the shortest-path distance in the device graph and $\lambda{=}0.3$ is chosen so a 2-hop pair receives $\sim$77\% of its native score, breaking ties toward native edges without eliminating non-native pairs from consideration when the native subgraph is disconnected. To strictly avoid SWAPs, we restrict the Kruskal pass to native edges of the device subgraph, falling back to non-native edges only if the native subgraph is disconnected on the chosen qubits. After the spanning tree we add the few remaining native edges that the importance ranking did not pick up, so the budget is slightly larger than $n{-}1$. The cost being avoided is concrete: on the 23-qubit heavy-hex subgraph used for QPU runs, the topology-agnostic Informed-$k$ pairs place $\sim$5--7 selected couplings on non-native edges with $d_G \ge 2$, each of which injects one SWAP per $p$-layer and inflates the native two-qubit count by $\sim$15--20\%. Restricting Kruskal to $\mathcal{E}_G$ eliminates this overhead entirely; Sec.~\ref{sec:hw-scale} reports zero inserted SWAPs across all three QPU runs. All simulator benchmarks in the Results section use the topology-agnostic Informed-$k$ selection; the hardware-aware variant is reserved for the QPU experiments (Sec.~\ref{sec:hw-scale}, Fig.~\ref{fig:ansatz_3seq_hh}).

\begin{figure*}[t]
  \centering
  \includegraphics[width=0.85\textwidth]{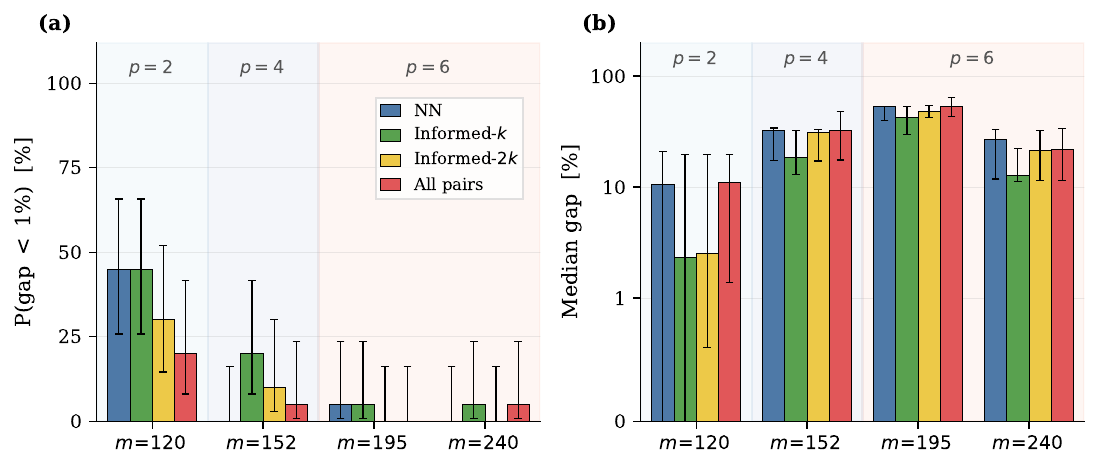}
  \caption{Ansatz-topology comparison across four benchmark sequences at PAGD-$K_{10}$ (layerwise parameterization, 20 seeds per topology). Each sequence is run at its designated depth $p$ (Table~\ref{tab:benchmarks}); the shaded background separates the $p{=}2$, $p{=}4$, and $p{=}6$ regions. Four topologies are compared: nearest-neighbor (NN), problem-informed top-$k$ pairs (Informed-$k$), top-$2k$ pairs (Informed-$2k$), and all $\binom{n}{2}$ pairs (All pairs). Panel~(a): P(gap~$<$~1\%) with Wilson 95\% CI. Panel~(b): median optimality gap with interquartile range (IQR; symlog $y$-axis).}
  \label{fig:ansatz_topology}
\end{figure*}

\subsection{Training Loss: From Ising Space to QUBO Space}\label{sec:loss}

PCE operates in a relaxed continuous space: the parameterized circuit produces expectation values (EVs) $\mathbf{e}(\boldsymbol{\theta}) \in [-1,+1]^m$, and a continuous loss function drives the optimizer toward configurations that, once decoded, yield good binary solutions. The choice of loss function is therefore critical, especially for constrained problems.

\paragraph{Standard Ising-space losses}
Prior PCE work~\cite{sciorilli2025pce} trains on the Ising energy of softened expectation values. Given the Ising decomposition of the normalized QUBO, $E_{\mathrm{Ising}}(\mathbf{s}) = \sum_{i<j} J_{ij}\,s_i s_j + \sum_i h_i\,s_i$, the \emph{tanh} loss substitutes $s_i \to z_i = \tanh(\alpha\, e_i)$ with $\alpha > 0$ and an optional saturation regularizer. Variants such as the Continuous Relaxation Annealing (CRA)-inspired loss~\cite{ichikawa2024controlling} use raw expectation values with a discretization penalty.

\paragraph{The linear-bias problem on dense-constraint QUBOs}
Ising-space losses share a structural vulnerability on dense-constraint problems. The QUBO penalty $P\,x_i x_j$ maps to the Ising term $\tfrac{P}{4}(1 - s_i - s_j + s_i s_j)$, and the $-\tfrac{P}{4}\,s_i$ contributions accumulate in the linear bias $h_i$. When the per-variable constraint count is large, $|h_i|$ dominates the quadratic couplings $J_{ij}$ after standard $Q/\max|Q|$ normalization, biasing the optimizer toward the trivial all-zero solution.

\paragraph{QUBO-space sigmoid loss}
We propose training directly in the QUBO space. Define the soft binary variables $\tilde{x}_i = \sigma(-\alpha\, e_i)$, where $\sigma$ denotes the logistic sigmoid. The training loss is then
\begin{equation}
\mathcal{L}_{\mathrm{QUBO}} = \tilde{\mathbf{x}}^\top Q\, \tilde{\mathbf{x}},
\label{eq:qubo-loss}
\end{equation}
where $Q$ is the original (un-normalized) QUBO matrix including penalty terms. The key difference is that penalty contributions activate only when both $\tilde x_i$ and $\tilde x_j$ are simultaneously near 1: the penalty gradient $\partial\mathcal{L}/\partial e_i \propto \sum_j Q_{ij}^{\mathrm{pen}}\,\tilde x_j$ vanishes when conflicting partners are inactive, so the penalty is \emph{conditional} rather than a constant linear bias. This lets the optimizer explore energy-favorable activations early, with EVs shaped by the competition between energy gain and constraint cost (Fig.~\ref{fig:decoder_comparison}).

\begin{figure*}[t]
  \centering
  \includegraphics[width=\textwidth]{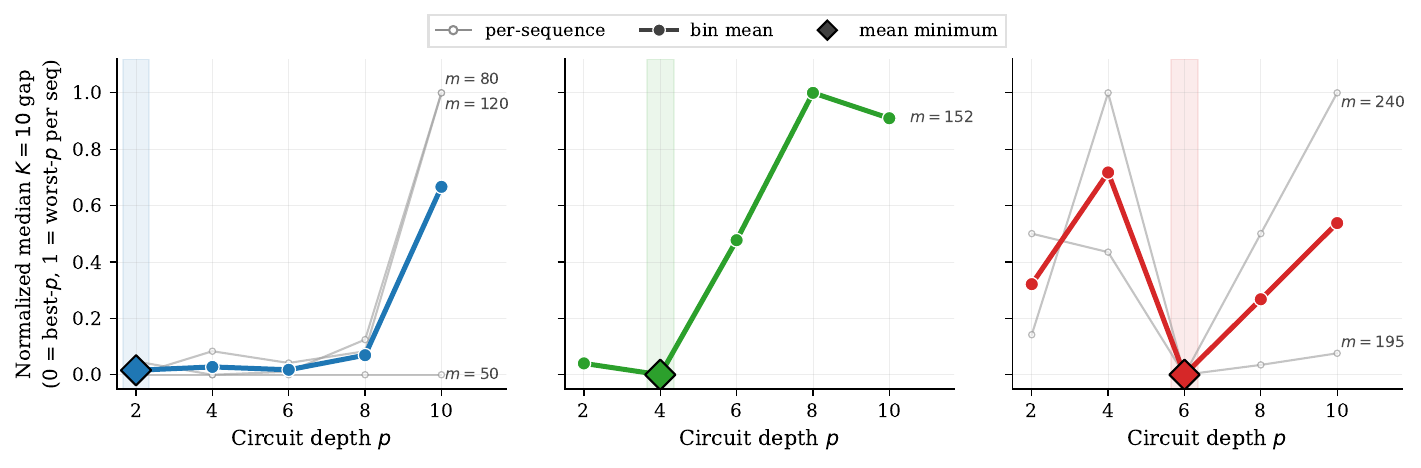}
  \caption{Circuit-depth $p$ selection across the six benchmark sequences (PCE, Informed-$k$ ansatz, layerwise parameterization, 20 seeds per $(m,p)$ configuration). Each subplot corresponds to one size class and shows the normalized median PAGD-$K_{10}$ decoded gap as a function of $p$; the normalization rescales each sequence's curve to $[0,1]$ by its own range so that the best-performing $p$ sits at~0 and the worst at~1. Gray lines show individual sequences (annotated by $m$ at the right edge), colored lines show the per-bin mean, and colored diamonds mark the minimum of the mean curve. Shaded vertical bands highlight the $p$ value adopted for each size class in this work. The mean minimum lies on the shaded band in every bin, and the optimum shifts from $p{=}2$ at small $m$ to $p{=}6$ at $m \in \{195, 240\}$ on the tested benchmarks, consistent with larger problems benefiting from greater circuit expressivity.}
  \label{fig:p_dependence}
\end{figure*}

\subsection{Decoding Schemes}\label{sec:decoders}
After quantum circuit optimization, PCE produces expectation values $e_i \in [-1, +1]$ for each encoded variable. These continuous values must be decoded into binary solutions $\mathbf{x} \in \{0,1\}^m$ that minimize the QUBO objective while respecting problem constraints. We introduce a new decoder (PAGD) and benchmark it against several baselines.

\subsubsection{Problem-Aware Guided Decoder (PAGD)}

Existing PCE decoders use either the EVs alone (sign rounding, Bernoulli sampling~\cite{sciorilli2025pce}) or the QUBO structure alone (classical greedy~\cite{glover2018tutorial}). PAGD scores each candidate variable commitment by both the \emph{marginal QUBO energy reduction} and the \emph{trained EV prior}.

Given the current partial assignment $\mathbf{x}$ with active (uncommitted) variable set $\mathcal{A}$, for each $i \in \mathcal{A}$ we compute the exact energy change from setting $x_i \gets 1$:
\begin{equation}
\Delta_i = Q_{ii} + 2\sum_{j:\, x_j=1} Q_{ij}.
\label{eq:pagd-marginal}
\end{equation}
This marginal includes the diagonal stem reward, stacking bonuses with already-selected stems (negative off-diagonal $Q_{ij}$), and constraint penalties with already-selected conflicting stems (positive $Q_{ij}$). A negative $\Delta_i$ means committing variable~$i$ reduces the QUBO energy.

The trained EV prior reuses the soft binary variable $\tilde{x}_i = \sigma(-\alpha\, e_i) \in [0,1]$ from Eq.~\eqref{eq:qubo-loss}, which is high when the trained circuit indicates $x_i = 1$ (i.e., $e_i < 0$). The combined score is
\begin{equation}
\text{score}_i = (-\Delta_i) \cdot \tilde{x}_i^{\,\beta},
\label{eq:pagd-score}
\end{equation}
where $\beta \geq 0$ controls the prior strength ($\beta = 0$ recovers pure classical greedy; $\beta = 1$ gives equal weighting). At each step, the variable $i^* = \argmax_i \text{score}_i$ with $\Delta_{i^*} < 0$ is committed, and all constraint-conflicting variables are pruned from $\mathcal{A}$. The process repeats until no energy-reducing commitment remains. The greedy commit-and-prune skeleton mirrors classical maximum-weight independent-set heuristics on the QUBO conflict graph~\cite{halldorsson1997greedy}; PAGD's contribution is the trained-prior weighting $\tilde x_i^{\beta}$ that re-orders ties in favor of variables the circuit is already biased toward.

\begin{figure*}[t]
  \centering
  \includegraphics[width=\textwidth]{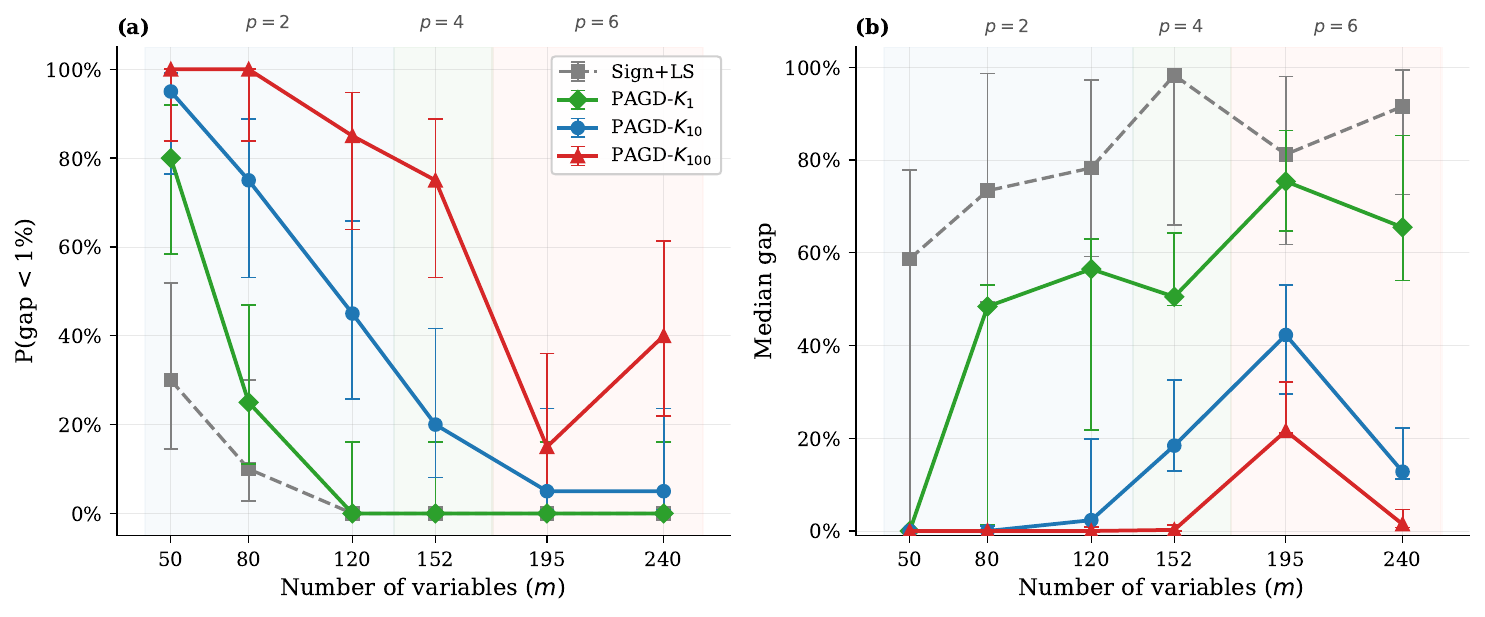}
  \caption{Decoder comparison across six benchmark sequences (PCE, layerwise parameterization, Informed-$k$, 20 seeds).
  Each sequence is evaluated at the circuit depth $p$ that minimizes median gap for that problem size:
  $p{=}2$ for $m{\leq}120$, $p{=}4$ for $m{=}152$, $p{=}6$ for $m \in \{195, 240\}$.
  (a)~Near-optimal recovery rate P(gap$<$1\%) with Wilson 95\% CI.
  (b)~Median optimality gap with IQR.
  Sign+LS: greedy sign-decode followed by best-improve local search.
  PAGD: $\sigma_{\text{noise}}{=}0.2$, $\alpha{=}8$, $\beta{=}1.0$; $K$ restarts per decode.
  Ground truths verified by exact integer programming.
  }
  \label{fig:decoder_comparison}
\end{figure*}

\paragraph{PAGD-$K$ variant} To escape greedy local optima, we run PAGD $K$ times with small Gaussian perturbations added to the EV vector before computing the prior, $e_i^{(k)} = e_i + \eta_i^{(k)}$ with $\eta_i^{(k)} \sim \mathcal{N}(0, \sigma_{\text{noise}}^2)$ drawn independently per restart $k$ and per variable $i$. The run with lowest QUBO energy is returned. The restart count $K$ scales with problem size: empirically, $K_{10}$ saturates performance at $m \leq 80$ (95--100\% P($<$1\%)), $K_{100}$ achieves 75--100\% at $m \leq 152$, and larger instances ($m \geq 195$) remain restart-limited even at $K_{200}$ (Figs.~\ref{fig:decoder_comparison}, \ref{fig:ksweep}). The total complexity is $O(K \cdot w \cdot m)$ where $w = |\{i : x_i = 1\}|$ is the Hamming weight of the output (typically 5--17 for our mRNA instances), giving effectively $O(Km)$ per decode. The marginal Eq.~\eqref{eq:pagd-marginal} also exploits mRNA stacking bonuses by construction: committing a stem makes its stacking partners more attractive on the next step, producing a chain effect along stacked stems.



\subsubsection{Baseline Decoders}
\label{sec:baseline-decoders}
We compare PAGD against two classical baselines that represent the standard repertoire of PCE decoding approaches.

\paragraph{Sign Rounding}
The simplest decoder sets $x_i = \mathbb{I}[e_i < 0]$ (negative EVs map to $x_i{=}1$, non-negative to $x_i{=}0$)~\cite{sciorilli2025pce}, in $O(m)$ time and using no QUBO structure beyond the EVs. On the dense-constraint mRNA QUBO (optimal Hamming weight only 5--20\% of $m$), it produces frequent constraint violations by committing high-weight variables simultaneously without feasibility checking. We also tested the Bernoulli-sampling decoder of Ref.~\cite{sciorilli2025pce} (which draws bitstrings from $\Pr(x_i{=}1) = \sigma(-\alpha\, e_i)$); at matched classical-iteration budget it offered no measurable advantage over sign rounding on these dense-constraint instances and is omitted from the comparison.

\paragraph{Sign + Local Search (Sign+LS)}
Applies sign rounding followed by up to $T{=}3$ passes of greedy single-bit-flip local search~\cite{glover2018tutorial, sciorilli2025competitive}: at each pass, every uncommitted bit is flipped if doing so reduces the QUBO energy while leaving all constraints satisfied. Complexity is $O(Tm^2)$. Sign+LS repairs some constraint violations and reduces energy post-rounding, making it a stronger baseline than sign rounding alone. It is included as the best purely-classical reference point for decoding from PCE expectation values without QUBO-structure information in the scoring step.

\subsection{Experimental Setup}\label{sec:experimental-setup}

\paragraph{Simulation}
A noiseless state-vector simulator computes the Pauli-correlator EVs for the six benchmarks (Table~\ref{tab:benchmarks}, \texttt{seq\_$m$} naming throughout) using Informed-$k$ topology and the layerwise parameterization ($2p{+}1$ parameters). For each sequence we swept $p \in \{2,4,6,8,10\}$ and four topologies (NN, Informed-$k$, Informed-$2k$, All); the per-size best depth (median PAGD-$K_{10}$ gap under Informed-$k$) is $p{=}2$ for $m \leq 120$, $p{=}4$ for $m{=}152$, $p{=}6$ for $m \in \{195,240\}$. Training uses COBYLA~\cite{powell1994cobyla} on Eq.~\eqref{eq:qubo-loss} with $\alpha{=}8$, 20 random seeds, 50--160 iterations. The three hardware-scale instances use the same pipeline at $p{=}10$.

\paragraph{Hardware}
QPU experiments use the IBM Heron processors \texttt{ibm\_pittsburgh} and \texttt{ibm\_aachen}. For each of the three hardware-scale sequences we extract a 23-qubit, 24-edge native subgraph of the target device containing two closed hexagonal cells, and optimize the physical-to-logical qubit relabeling by simulated annealing to maximize the total QUBO importance carried by the 24 selected edges (Sec.~\ref{sec:informed-ansatz}, Fig.~\ref{fig:ansatz_3seq_hh}). The resulting ansatz captures $96.7\%$ of the topology-agnostic top-24 importance ceiling, vs.\ $83.5\%$ for the bare maximum-importance spanning tree on the same 23 qubits. All 24 entangling pairs are native, so transpilation introduces no SWAP gates. Edge coloring partitions the 24 pairs into 3 parallel sublayers of 8 gates each, halving the per-block MS depth from 24 sequential gates to 3. We retain the layerwise parameterization of the simulator benchmarks ($2p{+}1$ COBYLA parameters: one $R_y$ angle per layer, one MS angle per layer shared across all 24 pairs), so the same 21-parameter scheme drives both the simulator and QPU pipelines at $p{=}10$. Each MS gate decomposes into 2 native two-qubit gates, so every circuit transpiles to 480 native CZ gates at depth 256, SWAP-free. Circuits are transpiled with Qiskit Primitive V2 at optimization level~3 with XX Dynamical Decoupling~\cite{viola1998dynamical} and Twirled Readout Error eXtinction (T-REX)~\cite{van_den_Berg_2022} (EstimatorV2 \texttt{resilience\_level}{=}1), and trained with COBYLA at $2^{14}$ shots per evaluation. Each COBYLA iteration on \texttt{ibm\_pittsburgh} and \texttt{ibm\_aachen} took approximately 25~seconds at the $2^{14}$-shot, three-Pauli-setting budget, so a 100-iteration QPU run completed in $\sim$40~minutes of QPU time per sequence, executed within a single device-calibration window to ensure consistent fidelities across the run.

\paragraph{Reproducibility}
The benchmark sequences were generated as random mRNA strings and converted to QUBO instances using the formulation of Ref.~\cite{alevras2024mrna} as described in section ~\ref{sec:QUBO}. Ground-truth optimal bitstrings for all six simulation benchmarks and the three hardware-scale instances were computed and verified with the CPLEX integer programming solver on the full QUBO. PAGD uses fixed hyperparameters $\alpha{=}8$, $\beta{=}1.0$, $\sigma_{\text{noise}}{=}0.2$ throughout, with no per-instance tuning.

\section{Results}
\label{sec:results}

\begin{figure*}[!b]                                                                                                                         
\centering                             
\includegraphics[width=\textwidth]{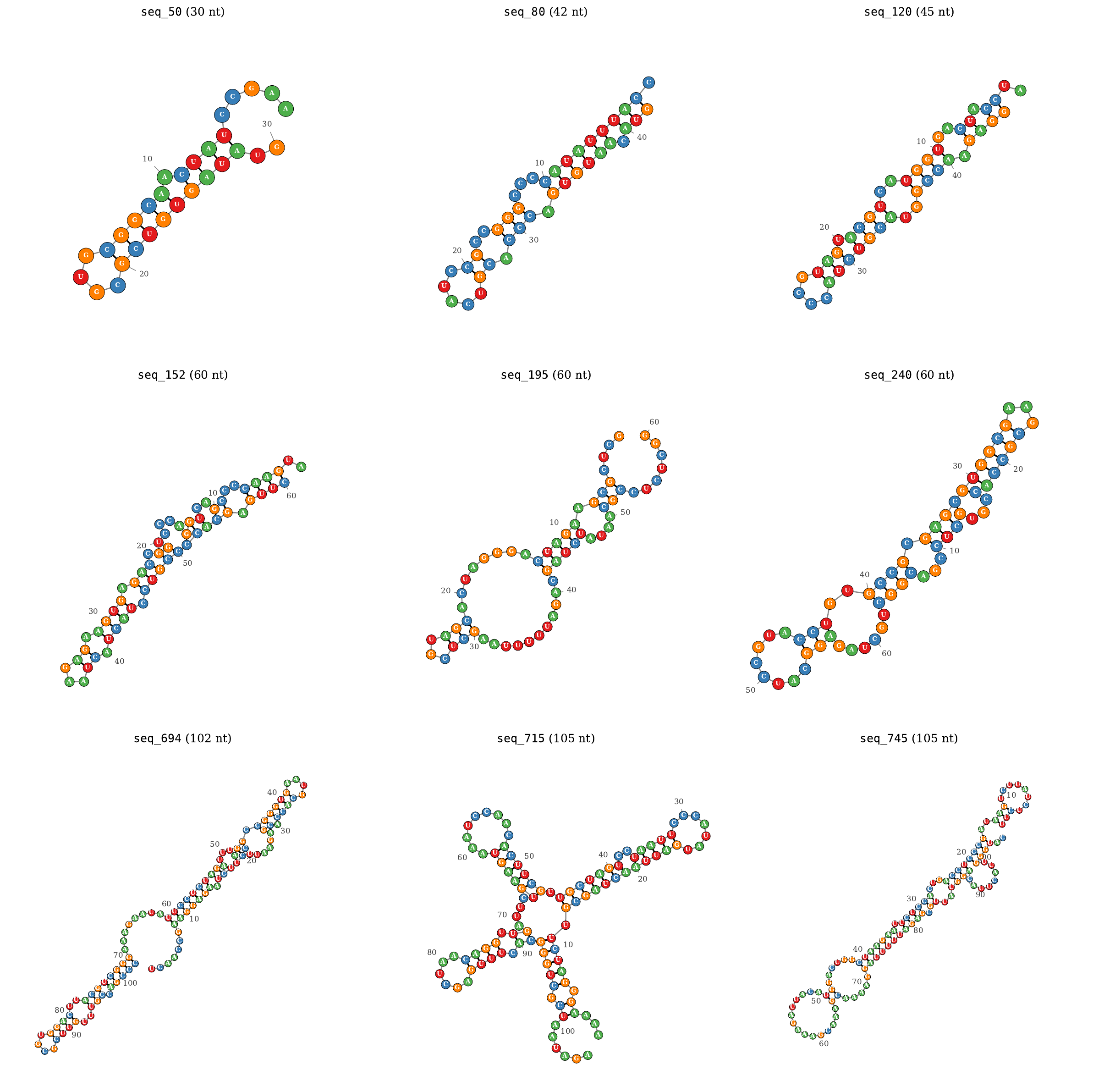}
\caption{Lowest-energy mRNA secondary-structure folds for all nine benchmark sequences. All nine sequences are evaluated in simulation. Top two rows: $m\in\{50,80,120,152,195,240\}$. Bottom row: $m\in\{694,715,745\}$ ($\sim$100~nt each); these three are additionally deployed on IBM Heron QPUs. Each panel is annotated with the sequence identifier and nucleotide length; per-base coloring follows the convention A=green, U=red, G=orange, C=blue. All folds shown are CPLEX-verified optima of the full QUBO.}                                              
\label{fig:rna_structures_main}        
\end{figure*}

Fig.~\ref{fig:rna_structures_main} shows secondary-structure renderings of all nine benchmark mRNA sequences in their CPLEX-verified optimal folds: all nine benchmark sequences are evaluated in simulation (top two rows, $m\in\{50,80,120,152,195,240\}$; bottom row, $m\in\{694,715,745\}$); the three sequences in the bottom row are additionally deployed on the QPU. Per-base coloring follows the convention A=green, U=red, G=orange, C=blue. The density and topology of stem--loop pairings translate directly into the off-diagonal coupling structure of the QUBO matrix~$Q$ and, in turn, into the qubit-pair importance scores used by the Informed-$k$ ansatz selection.

For each seed, the PAGD-$K$ outcome is the best-of-$K$ decoded gap (the lowest gap across the $K$ restarts of that seed). We report two metrics on this per-seed best-of-$K$: near-optimal recovery P($<$1\%), the fraction of seeds whose best-of-$K$ gap falls below 1\% (with Wilson 95\% CI on the proportion); and the across-seed median best-of-$K$ gap with IQR. QUBO energies are in the natural units of $Q$.

\subsection{Ansatz Topology Under PAGD}\label{sec:topology}

Fig.~\ref{fig:ansatz_topology} compares the four entangling topologies at PAGD-$K_{10}$ on the four largest benchmarks ($m \in \{120, 152, 195, 240\}$). Two patterns emerge. First, the fully-connected \emph{All pairs} topology is the weakest choice at $m \in \{120,152,195\}$ ($\leq$20\% P($<$1\%) at $m{=}120$, $\leq$5\% at $m \in \{152,195\}$; median gaps 11--54\%), consistent with COBYLA struggling on the larger parameter count under the layerwise budget; at $m{=}240$ no topology achieves a sizeable P($<$1\%). Second, \emph{Informed-$k$} consistently minimizes the median decoded gap relative to NN, and matches or beats NN on P($<$1\%) at every size. The gain is largest where the QUBO connectivity is densest ($m \geq 152$), and we adopt Informed-$k$ as the default for all decoder comparisons below.

\subsection{Circuit-Depth Selection}\label{sec:pdep}

With the ansatz topology fixed to Informed-$k$, Fig.~\ref{fig:p_dependence} sets the per-sequence circuit depth used hereafter ($p{=}2$ for $m \in \{50,80,120\}$, $p{=}4$ for $m{=}152$, $p{=}6$ for $m \in \{195,240\}$). The per-bin minimum of the normalized $K_{10}$ gap matches the adopted $p$ in every size class, with the optimum shifting from $p{=}2$ at small $m$ to $p{=}6$ at $m \in \{195,240\}$, consistent with deeper circuits being needed at larger compression. Beyond $p{=}6$, $p \in \{8,10\}$ degrade for all six sequences, consistent with COBYLA struggling on larger parameter spaces.

\subsection{Decoder Comparison}

With the ansatz topology and circuit depth fixed, Fig.~\ref{fig:decoder_comparison} reports the headline decoder comparison: P($<$1\%) and median gap for Sign+LS and PAGD at $K \in \{1,10,100\}$, all under Informed-$k$ with 20 seeds and the best-$p$ per size.

Sign+LS reaches non-zero P($<$1\%) only at $m{=}50$ (30\%) and has 0\% for $m \geq 80$ (median gaps 58--98\%). PAGD-$K_1$, relying solely on the trained prior, already achieves 80\% / 25\% at $m{=}50/80$, but fails at $m \geq 120$, indicating that a single greedy commit order is insufficient at larger scales. With restarts, PAGD-$K_{100}$ achieves gap$<$1\% in 75--100\% of seeds for $m \leq 152$ and drops to 15--40\% at $m \in \{195,240\}$, pointing to a restart-budget bottleneck at the largest scales, dissected next.

\subsection{Restart Budget and the Value of Trained Priors}

\begin{figure}[t]
  \centering
  \includegraphics[width=\columnwidth]{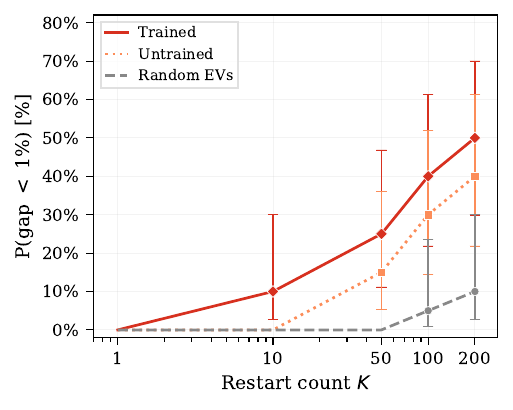}
  \caption{P(gap$<$1\%) versus restart count $K \in \{1,10,50,100,200\}$ on \texttt{seq\_240} ($m{=}240$, $n{=}14$ qubits, 20 seeds, layerwise, Informed-$k$, $p{=}6$). Three EV conditions: \emph{PCE-trained} (full COBYLA convergence); \emph{untrained} (random parameter initialization $\mathbf{x}_0 \sim \mathcal{U}[-\pi,\pi]$, zero COBYLA iterations: a single circuit evaluation at $\mathbf{x}_0$); and \emph{random EVs} $\sim \mathcal{U}[-1,+1]$. The \emph{untrained} condition is the strictest control: it shares the exact circuit, encoding, and noise model with the trained case, so any gap separation is attributable to the trained parameters alone. Error bars: Wilson 95\% CI.}

  \label{fig:ksweep}
\end{figure}

To isolate component contributions at large scales, Fig.~\ref{fig:ksweep} compares three EV conditions on \texttt{seq\_240} across $K \in \{1,10,50,100,200\}$. Trained PCE improves steeply with $K$ (10\% P($<$1\%) at $K{=}10$, 50\% at $K{=}200$), beating the untrained circuit by $+$10~pp at $K{=}200$ and the random-EV baseline by $+$40~pp, confirming that PCE training encodes useful structure at this scale.

\subsection{Hardware-Scale Results}
\label{sec:hw-scale}

\begin{figure}[t]
\centering
\includegraphics[width=\linewidth]{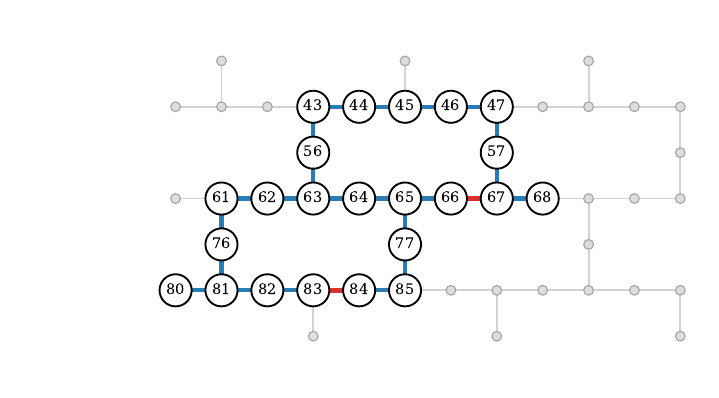}
\caption{Hardware-aware Informed-$k$ ansatz on the 23-qubit subgraph used for \texttt{seq\_745} on \texttt{ibm\_aachen}, drawn on a fragment of the surrounding heavy-hex device (gray). Node labels are the physical qubit indices on \texttt{ibm\_aachen}. Solid blue: the 22 native pairs of the QUBO-importance maximum spanning tree. Solid red: the 2 remaining native heavy-hex edges within the subgraph, added back to close the two hexagonal rings as described in Sec.~\ref{sec:informed-ansatz}. All 24 entangling pairs are native, so the circuit transpiles SWAP-free. The same edge structure is used for \texttt{seq\_694} and \texttt{seq\_715} on different physical qubit subsets.} 
\label{fig:ansatz_3seq_hh}
\end{figure}

We report combined simulator and QPU results on three $\sim$100-nucleotide hardware-scale instances: \texttt{seq\_694} (102~nt), \texttt{seq\_715}, \texttt{seq\_745} (both 105~nt). PCE encoding requires $n{=}23$ qubits for all three. The hardware-aware Informed-$k$ connectivity for \texttt{seq\_745}, constructed using the hardware-aware procedure described in Sec.~\ref{sec:informed-ansatz}, is shown in Fig.~\ref{fig:ansatz_3seq_hh}; all 24 selected pairs are native on the heavy-hex device, so the transpiled circuits contain no SWAP gates.

We train with the QUBO-space sigmoid loss, layerwise parameterization, and circuit depth $p{=}10$. The simulator runs use the AWS-deployed state-vector pipeline (Sec.~\ref{sec:experimental-setup}) at up to 800 COBYLA iterations averaged over 10 random seeds per sequence. The QPU runs are executed on \texttt{ibm\_pittsburgh} (\texttt{seq\_715}) and \texttt{ibm\_aachen} (\texttt{seq\_745}, \texttt{seq\_694}) with $2^{14}$ shots per evaluation, XX dynamical decoupling, and T-REX (resilience\_level=1); each QPU run is capped at 100 COBYLA iterations, well past the loss plateau but short of full convergence.

Table~\ref{tab:hw_vs_sim} summarizes the running-best PAGD gap-to-CPLEX-optimum at $K\in\{1, 10, 100, 200\}$ for all six (sequence, source) combinations.

\paragraph{Discussion}
The single QPU running-best per sequence is compared against the simulator distribution (per-seed best-of-$K$ on the same circuit and PAGD pipeline) as a one-sample falsification test, not a reproducibility claim. The QPU running-bests match or beat the simulator mean at $K \geq 100$ on all three sequences, and beat it at every $K$ reported on \texttt{seq\_694} and \texttt{seq\_745} (Table~\ref{tab:hw_vs_sim}): \texttt{seq\_694} (\texttt{ibm\_aachen}) recovers the CPLEX optimum exactly at $K_{200}$ (0.0\%) and reaches 7.4\% already at $K_{10}$; \texttt{seq\_715} (\texttt{ibm\_pittsburgh}) reaches 5.9\% at $K_{200}$ vs.\ simulator mean $14.4 \pm 4.2$\%; \texttt{seq\_745} (\texttt{ibm\_aachen}) reaches 18.0\% vs.\ simulator $32.3 \pm 4.2$\%. The trained PCE prior survives transit to noisy hardware at this scale; a multi-run QPU campaign yielding a hardware-side distribution is left to future work.

\begin{table*}[t]
\centering
\caption{Hardware-scale PAGD running-best gap-to-CPLEX-optimum (\%) at $K\in\{1,10,100,200\}$. Simulator: mean$\pm$1$\sigma$ across 10 seeds, with PAGD applied to the post-converged expectation-value vector of each seed. Hardware: running-best across 100 COBYLA iterations on IBM Heron. Lower is better.}
\label{tab:hw_vs_sim}
\small
\setlength{\tabcolsep}{4pt}
\renewcommand{\arraystretch}{1.15}
\begin{tabular}{lrrr|cccc|cccc}
\toprule
Sequence & $m$ & $L$ & $n$ & \multicolumn{4}{c|}{Simulator (mean over $N$ seeds)} & \multicolumn{4}{c}{Hardware (single run)} \\
         &     &     &     & $K{=}1$ & $K{=}10$ & $K{=}100$ & $K{=}200$               & $K{=}1$ & $K{=}10$ & $K{=}100$ & $K{=}200$ \\
\midrule
\texttt{seq\_694} & 694 & 102 & 23 & 65.7\,$\pm$\,14.6 & 44.5\,$\pm$\,11.6 & 18.6\,$\pm$\,8.0 & 15.4\,$\pm$\,8.6 & 33.2 & 7.4 & 7.3 & 0.0 \\
\texttt{seq\_715} & 715 & 105 & 23 & 56.5\,$\pm$\,17.8 & 29.8\,$\pm$\,8.5 & 16.9\,$\pm$\,3.0 & 14.4\,$\pm$\,4.2 & 20.2 & 12.5 & 6.5 & 5.9 \\
\texttt{seq\_745} & 745 & 105 & 23 & 72.5\,$\pm$\,11.0 & 49.2\,$\pm$\,6.7 & 39.7\,$\pm$\,5.9 & 32.3\,$\pm$\,4.2 & 30.5 & 30.0 & 18.5 & 18.0 \\
\bottomrule
\end{tabular}

\end{table*}

\section{Discussion}\label{sec:discussion}

The results above support three findings. (i) The trained PCE prior concentrates its advantage at low restart budgets ($K{=}1$--$10$), reducing the amount of classical multi-start decoding needed after a circuit evaluation, directly reducing QPU cost per problem. (ii) Informed-$k$ is the best-performing topology in the $m \in \{120,152,195,240\}$ range; the fully-connected \emph{All pairs} topology is the weakest at $m \in \{120,152,195\}$, indicating that a problem-adapted pair budget captures the entangling structure that matters without overloading the optimizer (Fig.~\ref{fig:ansatz_topology}). (iii) The pipeline reaches a few-percent gap on the best-performing $\sim$100-nt instance on QPU (Table~\ref{tab:hw_vs_sim}), confirming that the training encodes problem-specific information at problem sizes relevant to mRNA therapeutic design. The progressive increase in QPU $K_{200}$ gap across the three hardware-scale instances (0.0\% on \texttt{seq\_694}, 5.9\% on \texttt{seq\_715}, 18.0\% on \texttt{seq\_745}; Table~\ref{tab:hw_vs_sim}) tracks two structural features of the underlying QUBOs that increase monotonically over the same ordering: the optimal Hamming weight (20, 22, 23 active stems) and the conflict-pair constraint count ($1.72$, $1.83$, $1.93 \times 10^5$). Larger Hamming weights demand longer commit chains from PAGD, and more constraints prune the active set more aggressively at each step, both of which deepen the greedy-trap landscape that restarts must escape.

\paragraph{Limitations}
The trained-prior benefit is instance-dependent: it is largest at intermediate difficulty where classical greedy gets trapped but a good prior can escape, smaller on easy instances (e.g., \texttt{seq\_50} where PAGD-$K_1$ already reaches 80\%), and diminishes on the hardest instances ($m \geq 195$) where even $K_{200}$ is restart-limited. PAGD hyperparameters $(\alpha, \beta, \sigma_{\mathrm{noise}})$ are fixed at $(8, 1.0, 0.2)$; a sensitivity sweep is left to future work. We do not include an iso-time comparison against a tuned classical heuristic (SA, tabu, branch-and-cut); CPLEX is used as a verification oracle, not an iso-time competitor. The hardware-scale results are framed as a feasibility demonstration, not a claim of quantum advantage.

\section{Conclusions and Outlook}\label{sec:conclusions}

We have presented a systematic investigation of the training--decoding pipeline for Pauli Correlation Encoding applied to dense-constraint mRNA secondary structure QUBOs. Our key findings are:

\begin{enumerate}
\item We adopt the QUBO-space sigmoid loss Eq.~\eqref{eq:qubo-loss} as the natural training objective for QUBO-defined problems: it directly optimizes the same QUBO objective evaluated by the decoder, and we observed empirically that solution quality on the present mRNA instances is comparable to the Ising-space loss (full comparison out of scope here). We therefore present the QUBO-space formulation as the principled default aligned with the decoder rather than as an empirical winner.

\item PAGD combines marginal QUBO energy reduction with the trained EV prior, achieving 75--100\% near-optimal recovery, defined as P(gap$<$1\%), for $m \leq 152$ at $K_{100}$ (Fig.~\ref{fig:decoder_comparison}), vs.\ 0\% for Sign+LS at $m \geq 80$. At $K_1$ (no restarts) PAGD reaches 80\% at $m{=}50$ purely on the trained prior.

\item PCE training produces a meaningful prior: on \texttt{seq\_240}, trained PAGD reaches 50\% P(gap$<$1\%) at $K{=}200$, beating the untrained-circuit control by 10~pp and the random-EV baseline by 40~pp (Fig.~\ref{fig:ksweep}).

\item The pipeline extends to three hardware-scale instances ($m{=}694$--$745$, 102--105~nt, $n{=}23$): on \texttt{seq\_694} a single QPU run on \texttt{ibm\_aachen} with PAGD-$K_{200}$ recovers the CPLEX optimum exactly (0.0\% gap) and reaches 7.4\% at $K{=}10$ (Table~\ref{tab:hw_vs_sim}); the other two QPU runs reach 5.9\% (\texttt{seq\_715}) and 18.0\% (\texttt{seq\_745}) at $K_{200}$, both below their matched simulator means (Table~\ref{tab:hw_vs_sim}).
\end{enumerate}

More broadly, this work demonstrates that the performance of compact quantum encodings on constrained problems depends critically on the \emph{alignment} between the training loss, the problem structure, and the decoder. We expect this principle to extend to other compact encodings (e.g.\ Dicke-state encodings~\cite{bartschi2020grover}, qudit-based maps~\cite{wang2020qudits}) on constrained combinatorial problems.

\paragraph{Outlook}
Several directions for future work emerge from this study.
Extending the benchmark to longer mRNA sequences beyond the $\sim$745-variable / 105-nt instances reported here will test whether the restart-count bottleneck observed at these scales persists, and whether circuit noise on the QPU can substitute for PAGD restarts to escape local optima. Applying the same training--decoding pipeline to other dense-constraint problem classes (e.g.\ scheduling, graph coloring, resource allocation) is a natural next step.

\section*{Acknowledgments}
The authors thank Yashrajsinh Jadeja and Haining Lin for their contributions to the prediction of classical RNA secondary structures. During the preparation of this work the authors used GitHub Copilot Enterprise (with Sonnet 4.6, Opus 4.6, and Opus 4.7) and OpenAI Codex CLI (with GPT-4.4 and GPT-4.5) to accelerate code generation and refactoring across the training and decoding pipeline, automate figure-rendering scripts, and rapidly prototype and iterate on candidate training-loss formulations and decoder architectures that informed the final designs of the QUBO-space sigmoid loss and the Problem-Aware Guided Decoder. The research questions, design choices, and evaluation were author-driven; AI assistance was used under continuous human guidance, and all AI-assisted output was reviewed and edited by the authors, who take full responsibility for the final content.

\clearpage
\bibliographystyle{IEEEtran}
\bibliography{ref}

\end{document}